\documentstyle[12pt,aasms4,psfig]{article}
\newcommand{\etal}{et al.\  }
\newcommand{\grados}{^\circ}
\lefthead{Mart\'\i nez-Delgado \etal}
\righthead{Sagittarius}

\begin{document}

\title{Tidal streams in the Galactic halo: Evidence for the Sagittarius northern
stream or traces of a new nearby dwarf galaxy}

\author{David Mart\'\i nez-Delgado} 
\affil{Instituto de Astrof\'\i sica de Canarias,
E38205-La Laguna, Tenerife, Canary Islands, Spain}

\author{Antonio Aparicio} 
\affil{Instituto de Astrof\'\i sica de Canarias,
E38205-La Laguna, Tenerife, Canary Islands, Spain}
\affil{Departamento de Astrof\'\i sica, Universidad de La Laguna,
E38200 - La Laguna, Tenerife, Canary Islands, Spain}

\author{M. \'Angeles G\'omez-Flechoso} 
\affil{Geneva Observatory, Ch. des Maillettes 51, CH-1290 Sauverny,
Switzerland}

\author{Ricardo Carrera}
\affil{Instituto de Astrof\'\i sica de Canarias, E38205-La Laguna, Tenerife,
Canary Islands, Spain}

\begin{abstract}

Standard cosmology predicts that dwarfs were the first galaxies to be formed
in the Universe and that many of them merge afterwards to form bigger galaxies
such as the Milky Way. This process would have left behind traces such as tidal
debris or star streams in the outer halo. We report here the detection of a
very low density stellar system at $50\pm 10$ kpc from the Galactic center
that could be related to the merger process. It could form part of the
Sagittarius northern stream or, alternatively, could be the trace of a
hitherto unknown dwarf galaxy. 

The dwarf galaxy in Sagittarius, the closest satellite of the Milky Way, is
currently being tidally disrupted and is a ``living'' test for galaxy formation
theories. The system found here is $60\grados$ away from the center of the
Sagittarius galaxy. If it is really associated with this galaxy, 
it would confirm
predictions of dynamical interaction models indicating that tidal debris from
Sagittarius could extend along a stream completely enveloping the Milky Way
in a polar orbit.
\end{abstract}


\section{Introduction}

A result of standard cosmology is that dwarf protogalaxies are the first to
born as individual systems in the Universe. Afterwards, many of these merge to
form larger galaxies such as the Milky Way. The way in which this process
takes place has consequences for the present-day structure of the Milky
Way. The significant issues (White \& Frenk 1991) are how the merging
efficiency compares with the star formation efficiency in the protogalactic
fragments, and how the fragment merging and disruption time-scales compare
with the age of the Milky Way. If fragments are able to form stars before
merging, they will collapse non-dissipatively. If disruption was not complete,
Galactic precursors should be visible today as dwarf galaxy satellites or as
stellar streams within the Galactic halo.

Of particular relevance has been the discovery of the Sagittarius (Sgr) dwarf
galaxy (Ibata, Gilmore, \& Irwin 1994), a Milky Way satellite in an advanced
state of tidal disruption which provides a ``living'' test for tidal
interaction models and for galaxy formation theories. Since its discovery, it
has been the subject of several studies dealing with its structure (Mateo,
Olszewski, \& Morrison 1998), kinematics (Ibata \etal 1997), metallicity and
stellar age distribution (Layden \& Sarajedini 2000), as well the
modeling of the dynamics of its interaction with and disruption by the Milky Way
(Vel\'azquez \& White 1995; Ibata \& Lewis 1998; G\'omez-Flechoso, Fux, \& Martinet 1999; Helmi \& White 2000; Ibata \etal
2000b; Johnston \etal 1999). It was soon apparent that its extent was larger
than at first assumed, and dynamical models predict that the stream associated
with the galaxy should envelop the whole Milky Way in an almost polar orbit.

Recently, two teams of the Sloan Digital Sky Survey (SDSS), have presented
results from the first year commissioning data (Yanni \etal 2000; Ivezic
\etal 2000). They cover two long, narrow regions centered on the celestial
equator, one of length $87\grados$ and the other of length $60\grados$, both
$2.5\grados$ wide. In practice, this represents a slice through the Galactic
halo. Surprisingly, they find two clear $\sim45\grados$ long parallel strips of blue
A-type stars, with magnitudes $\simeq19$ and $\simeq21$. They could be
respectively formed by blue horizontal-branch and blue-straggler stars at
$\simeq45$ kpc from the Sun. While these two types of stars are expected to
appear associated in old stellar systems, the natural explanation is that the
strips are the trace of an old stream in the outer Galactic halo. Theoretical models of Sgr by Ibata \etal (2000a) and the discovery of a stream of carbon stars possibly associated with the Sgr tidal tail (Ibata \etal 2000b) indicate that the structure found in the SDSS can be explained by the northern stream of the Sgr dwarf galaxy.

The idea that this stream belongs to a tidally disrupted dwarf galaxy is very
tempting. The best known candidate is the Sgr dwarf. However, the possibility that the stream could be the trace of one or
several hitherto unknown, tidally disrupted, dwarf galaxies cannot be
rejected. In this letter, we report the detection of a very low density 
stellar system in this region that could be related to the aforementioned
disruption processes. 

\section{Observations and analysis}

To test further whether a distinct stellar system really exists in the
region and to characterize its properties, we have studied a
$35\times35(')^2$ field centered at $\alpha ={\rm 14^h44^m27^s}$ $\delta = -01\grados00'00''$ in
order to obtain a color--magnitude diagram (CMD) reaching down to the
turn-off, if any, of the stream stars. This field is within the slice
surveyed by the SDSS team and overlaps the remarkable RR Lyr clump 
detected around those coordinates at distance of $\simeq45$ kpc (Ivezic
\etal 2000). In addition, it is some $60\grados$ north of the center of the
Sgr dwarf and also within the area predicted by dynamical models for
the Sgr dwarf's orbit.

The main limitation of this study is the low stellar density of the
system in question, which results in foreground stars belonging to the Milky
Way being the dominant population in the CMD. In such conditions, it is quite
possible that the main sequence (MS) is the only detectable signature (if
any) of the system in the CMD. It is also necessary to estimate the
foreground star contamination. For this purpose, we have observed a control
field at $\alpha = {\rm 15^h14^m40^s}$ $\delta = +10\grados00'00''$. This field is some
$10\grados$ away both from the SDSS slice and from the Sgr dwarf's
orbit. This should be far enough to be free from stream stars and not to far
for the foreground star distribution to change significantly.

Observations  were made with Wide Field Camera at the prime focus of the Isaac Newton Telescope, at the Observatorio del Roque de los
Muchachos of the Instituto de Astrofísica de Canarias, on the island of La
Palma (Canary Islands, Spain) in 2000 June 9th. The observing conditions were photometric, with
seeing around 1.$\arcsec$5. Total integration times were 1900 s in $B$ and
1800 s in $R$ for the target field and 2600 s in $B$ and 2000 s in $R$ for the control field. Bias and flat corrections were done with IRAF. DAOPHOT and ALLSTAR (Stetson 1994) were then used to obtain the instrumental photometry of the stars. In total 120 measures of 50 standards from the list of Landolt (1992) were observed in each band to obtain the extiction for each night and the final transformation to the Johnson-Cousins standard system. The standard errors of the extinction are 0.006 mag in B and 0.004 in R while the zero-point standard
errors of the photometric transformation are 0.007 for B and 0.005 for R. The total errors in the zero-point of the photometry are therefore about 0.01 in both bands.

\section{Results}

Figure 1 shows the CMDs of the control (a) and target (b and c) fields. A
pseudo-$V$ magnitude, defined as
$''V''=(B+R)/2$, is used in the vertical scale. Figure 1{\it a} gives the distribution of foreground Milky Way 
stars. When compared with this, Figure 1{\it b} shows an denser strip at
$(B-r)\simeq0.8$, $23.5\leq ''V''\leq 22$. Its color and shape correspond to
what is expected for the upper MS of an old stellar population (see Figure
1{\it c\/}), 
with the turn-off at $''V''\simeq 22.6\pm0.3$. Figure 1{\it c} shows an
isochrone with such characteristics overplotted on the target field
CMD. Three variable star candidates have also been found in a similar way to
 Ivezic \etal (2000), which are marked
with squares in the figure. Their average magnitude is
$''V''=19.2$. Incidentally, this is about the magnitude of an RR Lyr
population associated to the MS of an old stellar population such as that
plotted in Figure 1{\it c}.

A test has been performed to check the statistical significance of the
proposed MS. Stars have been counted in the CMDs of the target and
control fields in a box defined by $22.4\leq ''V''\leq 22.9$, $0.7\leq
(B-R)\leq 1.0$ (we will call this the MS-box). Since the overall foreground
star densities of the target and control fields can be different, we have
also counted the stars in the rest of the CMDs, but avoiding the MS
region. In practice stars have been counted in the areas $18\leq ''V''\leq
21$, $(B-R)\leq 1.5$ and $18\leq ''V''\leq 22$, $1.5<(B-R)$ and then summed
up to sample the whole CMD resolved population (we will call this the
W-box). Artificial star tests have been carried out following the method
describe in Stetson \& Harris (1988) to
compute completeness factors as functions of magnitude and color, in particular
for the MS-box. The completeness-corrected number of stars is summarized in
Table 1. Errors have been estimated assuming Poissonian statistics for the
star counts. The target and control CMDs can be scaled to each other from the
W-box counts. Doing this, the net contribution of the proposed MS to the
target field MS-box is found to be $N_{\rm MS}=94\pm 17$ stars, indicating, with
a high statistical significance level ($>99$\%) that the feature is real and
encouraging us to conclude that we are observing a real stellar system,
differentiated from the surrounding halo population.

The distance to the system can be calculated from the magnitude of the
turn-off if its absolute magnitude and the interstellar reddening are
known. The latter is $E(B-V)=0.040\pm 0.001$ (Schlegel, Finkbeiner, \& Davis
1998). The former depends on the stellar population. Assuming that it is an
old system, we have used the ZVAR algorithm (Gallart \etal 1999) to compute a
synthetic stellar population model with stellar ages in the interval 10--15
Gyr and metallicity 1/20 of the  solar value (matching an old dwarf spheroidal
galaxy population). The
turn-off magnitude of the model is $M_{''V''}=3.9$ and the resulting distance
modulus, $(m-M)_0=18.6\pm 0.5$, which corresponds to a distance $d_0=52\pm
13$ kpc. To check the sensitivity of these values to the assumed stellar
population, we have computed a model for a younger population (stellar ages
in the interval 5--11 Gyr, matching that of the Sgr dwarf galaxy main
body; Layden \& Sarajedini 2000). The model turn-off magnitude is then
$M_{''V''}=3.5$, which introduces changes in the distance of 20\%. The errors
quoted above for $(m-M)_0$ and $d_0$ include the observational error in the
turn-off $ ''V''$ and the uncertainty arising from the stellar population
assumptions.

Alternatively, the distance can be estimated from the magnitude of variable
star candidates, assuming that they are RR Lyr type. For metallicity 1/20 of
solar,
the absolute magnitude of the RR Lyr is $M_{''V''}=0.6$ (Lee, Demarque, \&
Zinn 1990). The resulting distance modulus is $(m-M)_0=18.5\pm 0.5$, which
corresponds to a distance $d_0=50\pm 10$ kpc. Changing the metallicity by a
factor 5 changes the absolute magnitude only by 0.1 mag. The quoted errors
here take into account the variability amplitude of RR Lyraes. Within the 
errors,
this distance is compatible with the turn-off based estimate. Combining both,
the resulting distance is $d_0=51\pm 12$ kpc. This corresponds to a distance to
the Milky Way center of $d_{\rm MW}=46\pm 12$ kpc, and to the center of the
Sgr dwarf galaxy (considered to be the globular cluster M 54),
$d_{\rm Sgr-cen}=46\pm 12$ kpc.

Outputs of the computed synthetic stellar population model are the surface
magnitude, $\mu_{''V''}$, and surface stellar mass density, $\Sigma$. They
are $\mu_{''V''}=31.0\pm 0.2$, $\Sigma=(3.3\pm 0.6)\times 10^{-2}$
$M_\odot$ pc$^{-2}$.

\section{Discussion and conclusions}

Figure 2 shows the position in the sky of our target  and control fields,
together with the Sgr dwarf main body and its southern stream (Mateo \etal 1998). The possible tidal stream discovered by the SDSS team is located along the celestial equator (14 h $<$RA $< $ 15.3 h; Yanny \etal 2000). Although this SDSS scan strip is nearly perpendicular to the predicted orbit of Sgr, Ibata 
\etal(2000) identify this stream as being part of the  Sagittarius northern stream (see also Mart\'\i nez-Delgado, Aparicio \& G\'omez-Flechoso 2000). This result suggests that the stellar system reported here could be  tidal debris from Sgr, situated  some $60\grados$ north of its center.

In order to test this hypothesis, we have run $N$-body simulations
of the Sgr plus  Milky Way (MW) systems. The MW and Sgr models correspond to the
Galaxy and s-B2c dwarf models described by G\'omez-Flechoso \etal (1999).
In these models, the Galaxy is a three component (halo + disk + 
bulge) galaxy and
the dwarf is described by a King model with mass $M_{\rm Sgr}=6.04 \times 10^7$ $M_{\odot}$, core radius $r_o=0.3$ kpc, central velocity dispersion $\sigma_o=15$ km s$^{-1}$ and tidal radius $r_{\rm t}=2.02$ kpc. The Sgr orbit has been selected to match
the observational constraints on its present position and velocity.
The apocenter and pericenter of the selected orbit are 60 and 15 kpc, and
the anomalistic period is $\sim 0.74$ Gyr. The galactocentric position
and velocity of the satellite model, after more than 4 Gyr orbiting
in the MW, are $(X,Y,Z)=(16.0,2.0,-5.9)$ kpc and 
$(U,V,W)=(241, -16, 248)$ km s$^{-1}$, satisfying the observational
data inside the observational limits. Other characteristics of 
Sgr's (internal velocity dispersion, surface density, etc.) are also 
fulfilled. We will refer to this snapshot of the simulation in the
following description.

The projection of the tidal tail model on the sky and its distance
to the Sun (as function of the RA) are shown in Figure 3 (gray dots). 
The corresponding observational data of Sgr 
(black square; Ibata \etal 1994) and its tidal tail (solid line; 
Mateo \etal 1998) are also plotted, as well as the potential 
detection by Majewski \etal (1999; white square). Other observations
related to the Sgr tidal tail are the globular cluster Pal 12
(cross; Dinescu \etal 2000) and the A-type star streams (Yanni \etal 2000) and  the RR Lyr star clump (Ivezic
\etal 2000)
(shaded areas) found in the SDSS\footnote{Errors in the distance of the overdensities have been calculated assuming $\pm 0.5$ and $\pm0.25$ mag. errors in the A-type  and RR Lyr star overdensities, respectively.}.
The model also fits the carbon stars stream reported by Ibata \etal (2000b),
although the uncertanties in the determination of distances to these stars make us to take this result with care. We remark that our model is consistent
with the observations of Sgr and its tidal tail and is 
also in good agreement with the unconfirmed detections of the Sgr tidal tail
in the SDSS.

Finally, our detection (black dot) can also be fitted by 
the Sgr model and, incidentally,  is close to our model predictions
for the apocentric distance of the orbit of the Sgr dwarf. Note that the apocenter and the northern stream are the densest parts of the tidal tail, being the favored regions to be detected. The position and
distance of our detection are also in good agreement with the theoretical models by different authors (Helmi \& White 2000; Ibata \etal 2000a; Johnston \etal 1999). This comparison between the putative Sgr detections and our model suggests that we have detected  tidal debris from the Sgr northern stream in the region close to the Sgr apocenter.

 However, the available observational data for Sgr
prevent us from reaching a definitive conclusion. Firstly, the narrow slice
 covered in the sky by the SDSS does not allow to determine the 
direction and the complete extension of the detected tidal stream. In addition, the lack of kinematic data  
is the main limitation in our comparison of the observational data with the 
Sgr models. Therefore, the possibility that the stream could be the trace of one or several hitherto unknown, tidally disrupted
 dwarf galaxies cannot be rejected. This would confirm the interesting possibility that we may detect more dwarf galaxies, or debris thereof, being currently "swallowed" by our Galaxy.

Summarizing, we can state that traces of a differentiated, very low density
stellar system have been detected in the region of the sky observed. The fact
that it is in an area where models predict the presence of the Sgr
northern stream supports the idea that we have in fact detected this stream
$60\grados$ north and $46\pm 12$ kpc away from the center of the Sgr
dwarf galaxy. However, the possibility that it corresponds to a hitherto
unknown galaxy, also probably tidally stripped, cannot be rejected. Further
observations to trace the stream spatially and kinematic data are needed to   determine definitively whether or not it is associated with the Sgr dwarf galaxy.

\newpage

\begin{deluxetable}{lcc}
\tablenum{1}
\tablewidth{400pt}
\tablecaption{Table 1. Star counts in the color--magnitude diagrams
\label{scounts}}
\tablehead{
\colhead{Field} & \colhead{MS-box} & \colhead{W-box} }
\startdata
Target     &        $187\pm14$  &  $1758\pm42$ \nl
Control    &        $ 88\pm 9$  &  $1580\pm41$ \nl
\enddata

\end{deluxetable}

\newpage

Figure 1. Color--magnitude diagrams of the control ({\it a\/}) and target 
({\it b\/} and {\it c\/})
fields. Panel {\it a}  provides the distribution of the foreground Milky Way
stars. The overdense strip at $(B-R)\simeq 0.8$, $22\leq ''V''\leq 23.5$ in
panel {\it b} CMD is interpreted as produced by a stellar system at a distance of
$51\pm 12$ kpc from us, which could make part of the Sagittarius northern
stream or, alternatively, could be the trace of a hitherto unknown tidally
disrupted dwarf galaxy. Squares represent variable star candidates. Panel {\it 
c}
shows the CMD of the target field with an old, low metallicity (age: 12 Gyr;
metallicity: 1/20 solar) isochrone from the Padua library (Bertelli \etal 1994) superposed. The
isochrone MS shape shows good agreement with the hypothetical target field 
MS. Also, the variable star candidates (squares) fall in the predicted region
of the horizontal branch.

Figure 2. Sky map showing the main body of the Sagittarius dwarf galaxy, the
southern stream mapping by Mateo \etal (1998) (small squares), and the positions of the target (double square) and
control (circled square) fields discussed in the present paper. The Milky Way plane is
represented by the gray band crossing the map. The SDSS slice coincides with
the celestial equator.

Figure 3. ({\it a\/}) Equatorial coordinates ($\alpha$, $\delta$) and ({\it b\/}) heliocentric
distances vs. $\alpha$ of our Sgr model (gray dots) and some observational
data that can be related with the Sgr dwarf galaxy (a detailed explanation is
given in the Sec. 4).

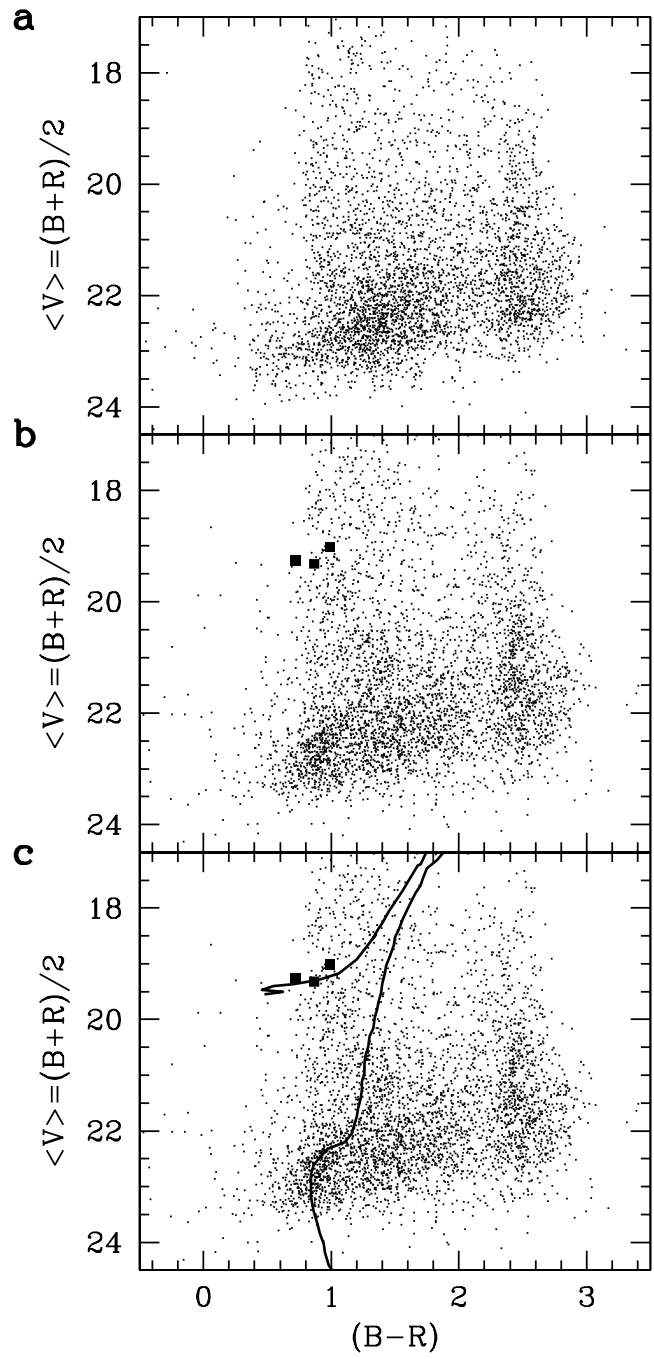
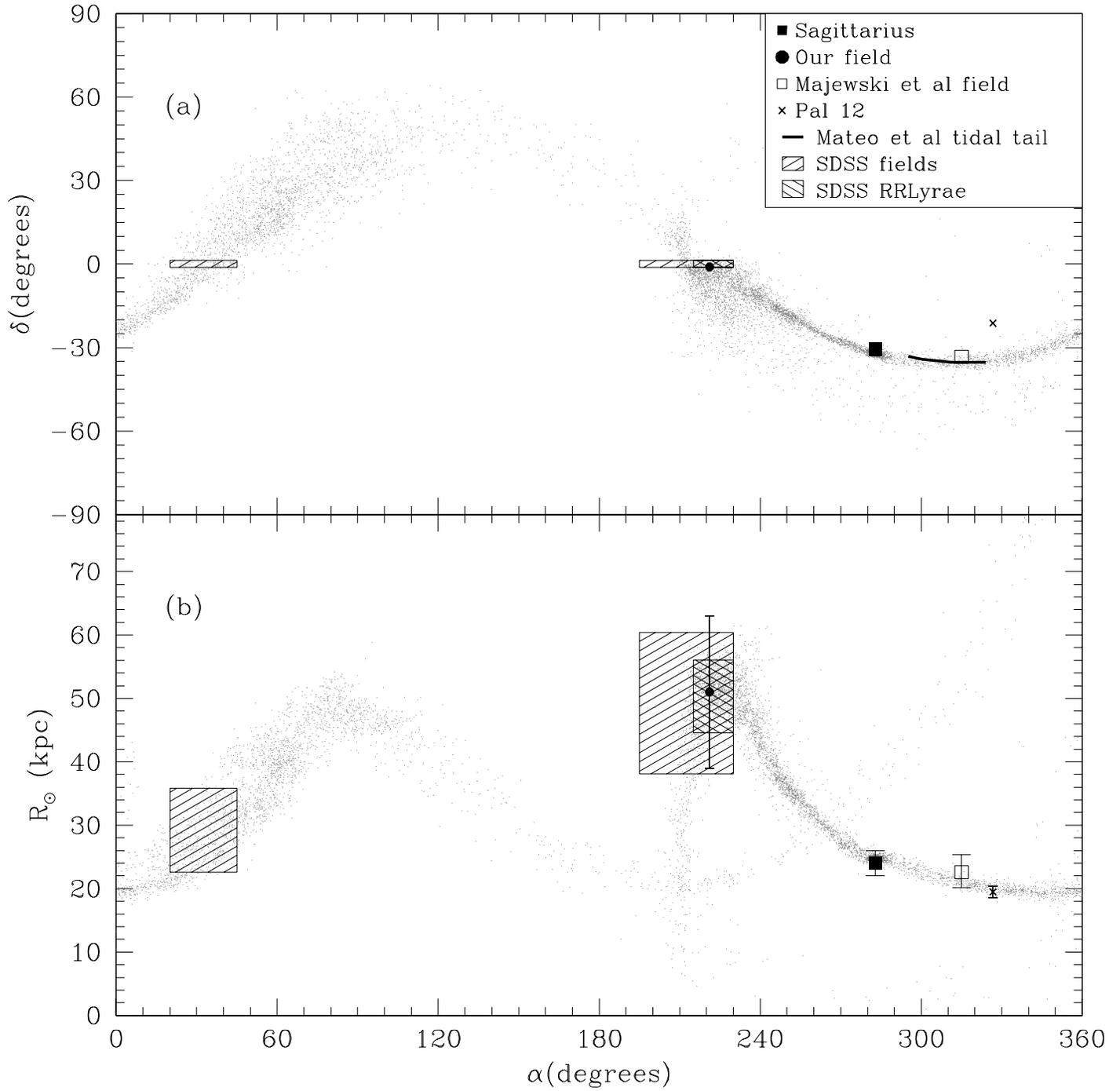
\end{document}